\documentclass[12pt,a4paper]{article}
\usepackage{amsmath}
\usepackage{amsfonts}
\usepackage{cite}

\renewcommand{\thesection}
  {\arabic{section}.\hspace{-.5em}}

\makeatletter
\renewcommand\section{
  \@startsection{section}{3}{\z@}%
  {-3.25ex\@plus -1ex \@minus -.2ex}%
  {1.5ex \@plus .2ex}%
  {\normalfont\normalsize\bfseries\mathversion{bold}}}
\renewcommand\subsection{
  \@startsection{subsection}{3}{\z@}%
  {-3.25ex\@plus -1ex \@minus -.2ex}%
  {1.5ex \@plus .2ex}%
  {\normalfont\normalsize\bfseries\mathversion{bold}}}
\makeatother

\makeatletter \@addtoreset{equation}{section} \makeatother
\renewcommand{\theequation}{\arabic{section}.\arabic{equation}}

\makeatletter
\renewcommand{\appendix}{
\renewcommand{\thesection}{\Alph{section}.\hspace{-.5em}}
\@addtoreset{equation}{section}
\renewcommand{\theequation}{\Alph{section}.\arabic{equation}}
\setcounter{section}{0}}
\makeatother

\newcommand{\Eqn}[1]{&\hspace{-0.5em}#1\hspace{-0.5em}&}
\newcommand{\nn}{\nonumber}
\renewcommand{\[}{\begin{equation}}
\renewcommand{\]}{\end{equation}}
\newcommand{\eqb}{\begin{eqnarray}}
\newcommand{\eqe}{\end{eqnarray}}

\newcommand{\bbZ}{{\mathbb Z}}

\newcommand{\grp}[1]{\mathrm{#1}}

\newcommand{\ttau}{\tilde{\tau}}
\newcommand{\iTheta}{\mathnormal{\Theta}}
\newcommand{\Rt}{R^\vee}
\newcommand{\mut}{\mu^\vee}
\newcommand{\bfR}{\boldsymbol{R}}
\newcommand{\calZ}{{\cal Z}}
\newcommand{\varth}{\vartheta}
\newcommand{\tj}{\tilde{\jmath}}

\begin{document}


\def\papertitlepage{\baselineskip 3.5ex \thispagestyle{empty}}
\def\preprinumber#1#2{\hfill
\begin{minipage}{1.0in}
#1 \par\noindent #2
\end{minipage}}

%
\papertitlepage
\setcounter{page}{0}
\preprinumber{YITP-12-15}{} 
\vskip 2ex
\vfill
\begin{center}
{\large\bf\mathversion{bold}
Seiberg--Witten prepotential for E-string theory\\[.5ex]
and random partitions
}
\end{center}
\vfill
\baselineskip=3.5ex
\begin{center}
  Kazuhiro Sakai\\

{\small
\vskip 6ex
{\it Yukawa Institute for Theoretical Physics, Kyoto University}\\[1ex]
{\it Kyoto 606-8502, Japan}\\
\vskip 1ex
{\tt ksakai@yukawa.kyoto-u.ac.jp}

}
\end{center}
\vfill
\baselineskip=3.5ex
\begin{center} {\bf Abstract} \end{center}

We find a Nekrasov-type expression for
the Seiberg--Witten prepotential for
the six-dimensional non-critical $E_8$ string theory
toroidally compactified down to four dimensions.
The prepotential represents the BPS partition function
of the $E_8$ strings wound around one of 
the circles of the toroidal compactification
with general winding numbers and momenta.
We show that our expression exhibits
expected modular properties.
In particular, we prove that it obeys
the modular anomaly equation
known to be satisfied by the prepotential.

\vfill
\noindent
March 2012


\setcounter{page}{0}
\newpage
\renewcommand{\thefootnote}{\arabic{footnote}}
\setcounter{footnote}{0}
\setcounter{section}{0}
\baselineskip = 3.5ex
\pagestyle{plain}
%

\section{Introduction}

An intriguing feature of local quantum field theories
in six dimensions is the existence of interacting
theories that involve self-dual tensor fields and strings.
The non-critical $E_8$ string theory, or the E-string theory,
is known as the simplest theory of this kind with (1,0) supersymmetry
\cite{Ganor:1996mu,Seiberg:1996vs,Klemm:1996hh,Ganor:1996pc,
      Minahan:1998vr}.
It includes one tensor multiplet
and possesses an $E_8$ global symmetry.
The theory was originally discovered
in the study of small $E_8$ instantons
in the $E_8\times E_8$ heterotic string theory compactified on K3
\cite{Ganor:1996mu,Seiberg:1996vs}.

While the whole picture of the theory
remains still mysterious,
toroidal compactification of the theory has been extensively studied.
Among others, compactification down to four dimensions is of
particular interest
\cite{Ganor:1996xd,Ganor:1996pc,Lerche:1996ni,
      Minahan:1997ch,Minahan:1997ct,Minahan:1998vr,
      Mohri:2001zz,Eguchi:2002fc}.
In this case,
the low energy effective theory admits a description in terms of
Seiberg--Witten theory
\cite{Seiberg:1994rs,Seiberg:1994aj}.
The Seiberg--Witten prepotential
represents the BPS partition function
of the E-strings wound around one of 
the circles of the toroidal compactification
with general winding numbers and momenta
\cite{Klemm:1996hh}.
Upon compactification
one can introduce eight Wilson line parameters
which break the $E_8$ global symmetry
\cite{Ganor:1996xd}.
The corresponding Seiberg--Witten curve was constructed
in the presence of the most general Wilson line parameters
\cite{Ganor:1996pc,Eguchi:2002fc}.
The prepotential can also be interpreted as the genus zero
topological string amplitude for the local $\frac{1}{2}$K3
or as the generating function of the partition functions
of ${\cal N}=4\ \grp{U}(n)$ topological Yang--Mills theories
on $\frac{1}{2}$K3
\cite{Minahan:1998vr}.

In this paper, we present an explicit expression
for the Seiberg--Witten 
prepotential
for the E-string theory.
We consider
the case with no Wilson line parameters.
Seiberg--Witten prepotential for this particular case
was studied in detail
by Minahan, Nemeschansky and Warner \cite{Minahan:1997ct}.
They considered the winding number expansion of the prepotential
and computed the expansion coefficients up to certain orders.
They elucidated that these coefficients are computed either
by solving the parametric relation among period integrals
or by recursively solving the modular anomaly equation.
Our expression is of Nekrasov type \cite{Nekrasov:2002qd}
and directly gives these coefficients at all orders.
The coefficients obtained from our expression are in perfect agreement
with those computed by the above methods
(verified for winding numbers $n\le 15$).
We show that our expression exhibits expected modular properties.
In particular, we prove that it satisfies
the modular anomaly equation of \cite{Minahan:1997ct}.

While the prepotential represents
the genus zero topological string amplitude
for the local $\tfrac{1}{2}$K3,
the sum over partitions in our expression
differs from the all-genus
topological string partition function for this Calabi--Yau threefold.
We clarify the difference of modular anomalies between them.

The organization of this paper is as follows.
In section~2, we recall some basic facts about
the Seiberg--Witten prepotential for the E-string theory
and briefly review how to compute its winding number expansion
by conventional methods.
In section~3, we present our Nekrasov-type expression
for the prepotential and discuss its structure.
In section~4, we show that our expression exhibits
expected modular properties. In particular, we prove that
it satisfies the modular anomaly equation.
We also make a comparison of modular anomalies between
our sum over partitions and
the all-genus topological string partition function
for the local $\frac{1}{2}$K3.
Section~5 is devoted to the discussion.
Some technical details are relegated to appendices.

\section{Seiberg--Witten prepotential for E-string theory}

In this section we recall some basic facts
about the Seiberg--Witten
prepotential for the E-string theory
and briefly review how to compute its winding number expansion
by two different methods
developed by Minahan, Nemeschansky and Warner
\cite{Minahan:1997ct}.

The E-string theory in six dimensions includes
one tensor multiplet. When the theory is toroidally compactified
down to four dimensions, the tensor multiplet turns into
a vector multiplet.
It contains a complex scalar field,
whose vev $\varphi$ parametrizes the Coulomb branch
of the vacuum moduli space of the compactified theory.
The low energy effective theory takes the form of
a four-dimensional ${\cal N}=2\ \grp{U}(1)$ gauge theory.
The effective action is fully
characterized by a prepotential $F_0$,
a holomorphic function of $\varphi$,
in the same way as
the original Seiberg--Witten theories
\cite{Seiberg:1994rs,Seiberg:1994aj}.
In the most general situation, the prepotential also depends on
the complex structure modulus $\tau$ of the torus
on which we compactify the theory
and eight Wilson line parameters $m_1,\ldots,m_8$
which break the $E_8$ global symmetry.
In this paper, we restrict ourselves to the case with 
no Wilson line parameters, namely the case of $m_1=\cdots=m_8=0$.

The prepotential represents the BPS partition function
of the E-strings wound around one of 
the circles of the toroidal compactification
with general winding numbers and momenta.
It can be expressed as
\[
F_0(\varphi,\tau)=
\sum_{n=1}^\infty\sum_{k=0}^\infty N_{n,k}
\sum_{m=1}^\infty\frac{p^{mn}q^{mk}}{m^3},
\]
where
\[
p:=e^{2\pi i\varphi},\qquad
q:=e^{2\pi i\tau}.
\]
Integer $N_{n,k}$ represents the multiplicity of
BPS states of winding number $n$ and momentum $k$.
The first few of them read \cite{Klemm:1996hh}
\begin{align}
\qquad\qquad&&N_{1,0}&=1,&\quad N_{1,1}&=252,&\quad N_{1,2}&= 5130,
&\quad\cdots&\qquad\nn\\
\qquad\qquad&&N_{2,0}&=0,&\quad N_{2,1}&=  0,&\quad N_{2,2}&=-9252,
&\quad\cdots&\ .\qquad
\end{align}
The prepotential can be viewed as
the genus zero topological string amplitude
for the local $\tfrac{1}{2}$K3.
The above integers are the 
numbers of rational curves in this Calabi--Yau threefold.
Indeed, these integers were computed in \cite{Klemm:1996hh}
by using the mirror map.

In practical computations, it is convenient to
express $F_0$ as a winding number expansion of the following form
\[\label{F0inZn}
F_0(\varphi,\tau)
= \sum_{n=1}^\infty Q^n Z_n(\tau),
\]
where
\[
Q:=q^{1/2}p.
\]
$Z_n$ is interpreted as
the partition function of topological ${\cal N}=4\ \grp{U}(n)$
Yang--Mills theory on $\frac{1}{2}$K3\cite{Minahan:1998vr}.
The first few of them read
\[\label{lowZn}
Z_1=\frac{E_4}{\eta^{12}},\qquad
Z_2=\frac{E_2E_4^2+2E_4E_6}{24\eta^{24}},\qquad
\cdots.
\]
Here $E_{2n}(\tau)$ are the Eisenstein series of weight $2n$
and $\eta(\tau)$ is the Dedekind eta function
(see Appendix~A).
For general $n$, $Z_n$ takes the form
\[\label{Gnform}
Z_n
=\frac{P_{6n-2}(E_2,E_4,E_6)}{\eta^{12n}},
\]
where $P_{6n-2}(E_2,E_4,E_6)$
denotes a polynomial in $E_2,E_4,E_6$
of weight $6n-2$.
The explicit form of $Z_n$ at low orders
can be calculated by either of the methods described below.

The prepotential can be computed from the
Seiberg--Witten curve of the theory.
In the present case, the Seiberg--Witten curve is given by
\[
y^2=4x^3-\frac{1}{12}E_4(\tau)u^4x-\frac{1}{216}E_6(\tau)u^6+4u^5.
\]
If we think of $u$ and $\tau$ as parameters,
it is an elliptic curve in the Weierstrass form.
An elliptic curve in this form
admits the following canonical parametrization
\[
y^2=4x^3-
\frac{1}{12}\frac{E_4(\ttau)}{\omega^4}x
-\frac{1}{216}\frac{E_6(\ttau)}{\omega^6}.
\]
Here $\ttau$ is the complex structure modulus
and $\omega$ (multiplied by $2\pi$)
is one of the fundamental periods
of the elliptic curve.
By comparing these two expressions, one can calculate
$\omega(u,\tau),\ttau(u,\tau)$ as series expansions in $1/u$.
They are related to the scalar vev $\varphi$
and the prepotential $F_0$ by
\eqb
\partial_u\varphi \Eqn{=} \frac{i}{2\pi}\omega,\\
\partial_\varphi^2 F_0 \Eqn{=}8\pi^3 i(\ttau-\tau).
\eqe
These relations parametrically determine the function
$F_0(\varphi,\tau)$.
The integration constants as well as
the normalizations of $F_0$ and $\varphi$
are determined so that $F_0$
gives the BPS partition function described above.
For further details of the calculation,
see \cite{Minahan:1997ct,Sakai:2011xg}.

An alternative way to compute the prepotential is to solve
the modular anomaly equation \cite{Minahan:1997ct}.
As is well known, $E_2(\tau)$ is not strictly a modular form,
but transforms anomalously.
The dependence of the prepotential on $E_2$
(i.e.~the modular anomaly of the prepotential)
is governed by the following equation
\[\label{MAEforF0}
\partial_{E_2}F_0
=\frac{1}{24}\left(
   \iTheta_QF_0\right)^2,
\]
where
$\iTheta_Q := Q\partial_Q = \tfrac{1}{2\pi i}\partial_\varphi$.
This modular anomaly equation was derived in \cite{Minahan:1997ct}
from the Seiberg--Witten description.
In terms of $Z_n$, the equation is written as
\[\label{MAEforZn}
\partial_{E_2}Z_n = \frac{1}{24}\sum_{k=1}^{n-1}k(n-k)Z_kZ_{n-k}.
\]
This equation recursively
determines $Z_n$ up to a piece which does not contain $E_2$.
Given the general structure (\ref{Gnform}),
the remaining ambiguity can be fixed by the gap condition
\[
q^{n/2}Z_n =\frac{1}{n^3}+{\cal O}(q^n).
\]
This condition follows
from the geometric structure of the local $\tfrac{1}{2}$K3
\cite{Minahan:1998vr}.

\section{Nekrasov-type expression}

In this section we present an explicit expression
for the Seiberg--Witten prepotential for the E-string
theory and discuss its structure.

Let $\bfR=(R_1,\ldots,R_N)$ denote an $N$-tuple of partitions.
Each partition
$R_k$ is a nonincreasing sequence of nonnegative integers
\[
R_k =\{
\mu_{k,1}\ge\mu_{k,2}\ge\cdots\ge\mu_{k,\ell(R_k)}>
\mu_{k,\ell(R_k)+1}=\mu_{k,\ell(R_k)+2}=\cdots=0\}.
\]
Here the number of nonzero $\mu_{k,i}$ is denoted by $\ell(R_k)$.
$R_k$ is represented by a Young tableau.
We let $|R_k|$ denote the size of $R_k$,
i.e.~the number of boxes in the Young tableau of $R_k$:
\[
|R_k|
:=\sum_{i=1}^\infty\mu_{k,i}
=\sum_{i=1}^{\ell(R_k)}\mu_{k,i}.
\]
Similarly,
the size of $\bfR$ is denoted by
\[
|\bfR|:=\sum_{k=1}^N|R_k|.
\]
We let 
$\Rt_k=\{\mu_{k,1}^\vee\ge\mu_{k,2}^\vee\ge\cdots\}$
denote the conjugate partition of $R_k$.
We also introduce the notation
\[
h_{k,l}(i,j):=\mu_{k,i}+\mu_{l,j}^\vee-i-j+1,
\]
which represents the relative hook-length of a box at $(i,j)$
between the Young tableaux of $R_k$ and $R_l$.

In our expression we consider
a sum over four partitions.
For our present purpose, it is convenient to
express these partitions as
\[
\bfR=(R_1,R_2,R_3,R_4)=(R_{11},R_{10},R_{00},R_{01}).
\]
The prepotential is then given by
\[\label{F0Estring}
F_0=(2\hbar^2\ln \calZ)\big|_{\hbar=0}\,,
\]
where
\eqb
\label{ZEstring}
\calZ\Eqn=
\sum_{\bfR}
Q^{|\bfR|}
\prod_{a,b,c,d}\,
\prod_{(i,j)\in R_{ab}}
\frac
{\varth_{ab}
 \left(\tfrac{1}{2\pi}(j-i)\hbar,\tau\right)^2}
{\varth_{1-|a-c|,1-|b-d|}
 \left(\tfrac{1}{2\pi}h_{ab,cd}(i,j)
       \hbar,\tau\right)^2}.
\eqe
Here the sum is taken over all possible
partitions $\bfR$
(including the empty partition).
Indices $a,b,c,d$ take values $0,1$,
while
a set of indices $(i,j)$ run over the coordinates of all boxes
in the Young tableau of $R_{ab}$.
$\varth_{ab}(z,\tau)$ are the Jacobi theta functions
(see Appendix~A).
$h_{ab,cd}(i,j)$ is the relative hook-length
defined between partitions $R_{ab}$ and $R_{cd}$.

We find that the above $F_0$
coincides with the prepotential
described in the last section.
Explicit forms of $Z_n$ obtained from this expression
are in perfect agreement with
those computed by either of the methods described in the last section
(verified for $n\le 15$).
We will also show in the next section that the above expression
exhibits expected modular properties.
In the rest of this section
let us make a few comments on the structure of our expression.

For any $\bfR$ with $R_{11}\ne\{0\}$,
the product in the sum vanishes.
This is because 
the Young tableau of $R_{11}\ne\{0\}$ always contains
a box at $(i,j)=(1,1)$,
where the theta function
in the numerator becomes $\varth_{11}(0,\tau)=0$.
Hence, $\calZ$ is actually a sum over three partitions
$(R_{10},R_{00},R_{01})$.
This structure is expected, as we consider the case with
no Wilson line parameters.
In the presence of the most general Wilson line parameters,
the BPS partition function of singly wound E-strings reads
$Z_1=\tfrac{1}{2}\eta^{-12}\sum_{a,b}\prod_{i=1}^8
 \varth_{ab}(m_i,\tau)$ \cite{Minahan:1998vr}.
By setting $m_i=0$, the product of $\varth_{11}$ vanishes
and the other three products add up to $Z_1=\eta^{-12}E_4$,
as is found in (\ref{lowZn}).
We expect that the expression with four partitions will be useful
when one considers the cases with nonzero Wilson line parameters.

The above $\calZ$ coincides with a special case of
the elliptic generalization of the Nekrasov partition function
for the $\grp{SU}(4)$ gauge theory
with 8 massless fundamental hypermultiplets
\cite{Nekrasov:2002qd,Hollowood:2003cv}.
More specifically, $\calZ$ can be expressed as
\[\label{ZSQCDlike}
\calZ
=\sum_{\bfR}
(-p)^{|\bfR|}
\prod_{k=1}^4
\prod_{(i,j)\in R_k}
\frac
{\varth_1\left(a_k+\tfrac{1}{2\pi}(j-i)\hbar,\tau\right)^8}
{\prod_{l=1}^{4}\varth_1
 \left(a_k-a_l+\tfrac{1}{2\pi}h_{kl}(i,j)\hbar,\tau\right)^2}
\]
with $a_k$ being set to half periods of the torus
\[
a_1=0,\qquad
a_2=\frac{1}{2},\qquad
a_3=-\frac{1+\tau}{2},\qquad
a_4=\frac{\tau}{2}.
\]
While physical implication of this coincidence is yet unclear,
it indicates that the analyticity of $\hbar^2\ln \calZ$
at $\hbar=0$ follows from that of
the Nekrasov partition function.
Therefore, $F_0$ given as in (\ref{F0Estring}) is indeed well defined.

\section{Modular properties and modular anomaly equation}

In this section we show that our expression exhibits
the modular properties described in section~2.
In particular, we derive the modular anomaly equation
(\ref{MAEforF0}) from our expression.

Recall that the Jacobi theta functions
can be expressed as
\eqb
\label{th1ins}
\varth_1\left(\frac{z}{2\pi},\tau\right)
 \Eqn{=}e^{-\frac{1}{24}E_2 z^2}\eta^3\sigma(z|2\pi,2\pi\tau),\\
\label{thkinc}
\varth_{k+1}\left(\frac{z}{2\pi},\tau\right)
 \Eqn{=}e^{-\frac{1}{24}E_2 z^2}\varth_{k+1}\sigma_k(z|2\pi,2\pi\tau),
\quad k=1,2,3.
\eqe
Here $\sigma$ and $\sigma_k$
are the Weierstrass sigma function and cosigma functions, respectively,
associated with the lattice
$2\pi\bbZ+2\pi\tau\bbZ$.
They are expanded as
\eqb
\label{wsexp}
\sigma(z|2\pi,2\pi\tau)
\Eqn{=}z-\frac{E_4}{2880}z^5-\frac{E_6}{181440}z^7+{\cal O}(z^9),\\
\label{wskexp}
\sigma_k(z|2\pi,2\pi\tau)
\Eqn{=}1-\frac{e_k}{2}z^2
 +\left(\frac{E_4}{576}-\frac{e_k^2}{8}\right)z^4+{\cal O}(z^6),
\eqe
where
\[\label{e1e2e3}
e_1=\frac{\varth_3^4+\varth_4^4}{12},\qquad
e_2=\frac{\varth_2^4-\varth_4^4}{12},\qquad
e_3=\frac{-\varth_2^4-\varth_3^4}{12}.
\]
The coefficients of these expansions can be computed
up to any order,
as explained in Appendix~B.
For the present purpose, it is enough to know that
the expansion coefficients of $\sigma$
are polynomials in $E_4,E_6$
while those of $\sigma_k$ are polynomials in $E_4,E_6,e_k$.

We are now in a position to
look into the modular properties of our expression.
As we saw in the last section,
(\ref{ZEstring}) is actually a sum over three partitions
$(R_{10},R_{00},R_{01})=(R_2,R_3,R_4)$.
Let us first
consider the contribution of the prefactors $\eta^3,\,\varth_{k+1}$
in (\ref{th1ins}), (\ref{thkinc}).
For each box in the Young tableau of $R_k$,
one obtains $\varth_k^8$ in the numerator as well as
$\eta^6\varth_2^2\varth_3^2\varth_4^2=4\eta^{12}$
in the denominator
after taking the product with respect to indices $c,d$.
From this structure it is clear that
the eta functions always appear in $\calZ$
through the combination $Q/\eta^{12}$.
The $\varth_k^8$ in the numerator can be expressed as
$\varth_2^8=16(e_2-e_3)^2$, $\varth_3^8=16(e_3-e_1)^2$,
$\varth_4^8=16(e_1-e_2)^2$.
Next, observe that the way theta functions
appear in $\calZ$ is entirely symmetric under the permutation
of $(\varth_2,\varth_3,\varth_4)$.
This can be seen easily by renaming the partitions
$(R_2,R_3,R_4)$ correspondingly.
Together with the expression of the theta functions
(\ref{thkinc}), (\ref{wskexp}),
this means that the way $e_1,e_2,e_3$ appear
in $F_0$ is also entirely symmetric.
It is also easy to see that
they appear in $F_0$
always as polynomials.
Any symmetric polynomial in $e_1,e_2,e_3$ is generated by
the elementary symmetric polynomials, which are identified as
\[
e_1+e_2+e_3=0,\qquad
e_1e_2+e_1e_3+e_2e_3=-\frac{1}{48}E_4,\qquad
e_1e_2e_3 =\frac{1}{864}E_6.
\]
Now we see that $\tau$ appears in $F_0$
only through polynomials in $E_2,E_4,E_6$
or through the combination $Q/\eta^{12}$.
Notice that modular weights are preserved in
(\ref{th1ins})--(\ref{wskexp})
if one assigns weight $-1$ to the expansion variable $z$.
As $\calZ$ is manifestly of weight~$0$,
it is obvious that $F_0$ is of weight $-2$.
Hence, we have shown that our expression indeed reproduces
the structure of $Z_n$ as in (\ref{Gnform}).

Next let us consider the modular anomaly.
The modular anomalies appear through $E_2$.
An important feature of the expressions
(\ref{th1ins})--(\ref{wskexp}) is that
$E_2$ appears only through the exponential prefactors
in (\ref{th1ins}), (\ref{thkinc}).
After evaluating all the products in (\ref{ZEstring}),
the total exponential factor becomes
\[
\exp\left(\frac{1}{12}|\bfR|^2E_2\hbar^2\right)
\]
for partitions $\bfR$.
Here we have used the following combinatorial identity
\[\label{keyidentity}
\sum_{k,l=1}^N
\sum_{(i,j)\in R_k}
\left(h_{kl}(i,j)^2-(j-i)^2\right)
=|\bfR|^2.
\]
We present a proof of this identity in Appendix~C.
It is now clear that $\calZ$
satisfies the following modular anomaly equation
\[\label{MAEforZ}
\partial_{E_2}\calZ=\frac{1}{12}\hbar^2\iTheta_Q^2 \calZ.
\]
By substituting
\[\label{lnZexp}
\calZ=\exp\left(
  \frac{1}{2}F_0\hbar^{-2}
  +{\cal O}(\hbar^0)\right),
\]
we obtain the modular anomaly equation (\ref{MAEforF0}).

Since our $F_0$ coincides with the genus zero topological
string amplitude for the local $\frac{1}{2}$K3
(evaluated at $m_i=0$),
one may expect that higher order coefficients
of the expansion of $\ln \calZ$ in $\hbar$
give higher genus topological string amplitudes.
However, this is not the case.
It is known that the all-genus topological string partition function
for the local $\frac{1}{2}$K3
\[\label{ZRESexp}
\calZ^{\mbox{\scriptsize $\frac{1}{2}$K3}}
=\exp\left(\sum_{g=0}^\infty
\hbar^{2g-2}
F_g^{\mbox{\scriptsize $\frac{1}{2}$K3}}
\right)
\]
satisfies the holomorphic anomaly equation \cite{Hosono:1999qc}
\[\label{HAEforZRES}
\partial_{E_2}\calZ^{\mbox{\scriptsize $\frac{1}{2}$K3}}
=\frac{1}{24}\hbar^2\iTheta_Q(\iTheta_Q+1)
\calZ^{\mbox{\scriptsize $\frac{1}{2}$K3}}.
\]
Despite the apparent difference,
equations (\ref{MAEforZ}) and (\ref{HAEforZRES})
give the same modular anomaly equation (\ref{MAEforF0})
for $F_0=F_0^{\mbox{\scriptsize $\frac{1}{2}$K3}}\big|_{m_i=0}$\,.
This coincidence, however, does not persist at $g\ge 1$.
Higher order parts of the expansion (\ref{lnZexp})
do not seem to have an immediate connection with
$F_g^{\mbox{\scriptsize $\frac{1}{2}$K3}}$
at $g\ge 1$
found in
\cite{Mohri:2001zz,Hosono:2002xj,Sakai:2011xg}.

\section{Discussion}

In this paper we have presented an explicit expression
for the Seiberg--Witten prepotential for 
the six-dimensional E-string theory
toroidally compactified down to four dimensions.
The expression is of Nekrasov type and
directly gives the coefficients of
the winding number expansion
of the prepotential at all orders.
We have shown that the expression exhibits
expected modular properties,
in particular we have proved that it satisfies
the correct modular anomaly equation.

We have pointed out that
the sum over partitions
in our expression,
denoted by $\calZ$, 
can be viewed as a special case of the elliptic generalization of
the Nekrasov partition function for the $\grp{SU}(4)$ gauge theory
with 8 fundamental hypermultiplets.
Currently, we do not have a good physical explanation of
this coincidence.
It would be of great interest if this could uncover
yet unknown dualities among field theories in six dimensions.

It is also mysterious that the prepotential is obtained as
the `genus zero part' of $\ln \calZ$ (up to an overall factor of 2),
where the sum $\calZ$ differs from the all-genus
topological string partition function for the local
$\tfrac{1}{2}$K3.
The local $\tfrac{1}{2}$K3 does not admit a local toric description
and no Nekrasov-type partition function
is known for such Calabi--Yau threefolds at present.
We hope our expression provides us with a new perspective on
the combinatorial study of topological string amplitudes
for non-toric Calabi--Yau threefolds.

\vspace{3ex}

\begin{center}
  {\bf Acknowledgments}
\end{center}

The author is the Yukawa Fellow and his work is supported
in part by Yukawa Memorial Foundation.
His work is also
supported in part by Grant-in-Aid
for Scientific Research from the Japan Ministry of Education, Culture, 
Sports, Science and Technology.

\vspace{3ex}

\newpage

\appendix

\section{Conventions of special functions}

The Jacobi theta functions are defined as
\[
\varth_{ab}(z,\tau)
:=\sum_{n\in \bbZ}\exp\left[
 \pi i\left(n+\frac{a}{2}\right)^2\tau
+2\pi i\left(n+\frac{a}{2}\right)\left(z+\frac{b}{2}\right)
\right],
\]
where $a,b$ take values $0,1$.
We also use the notation
\begin{align}
\hspace{3em}
&\varth_1(z,\tau):=-\varth_{11}(z,\tau),
&\varth_2(z,\tau):=\varth_{10}(z,\tau),&\hspace{5em}\nn\\
&\varth_3(z,\tau):=\varth_{00}(z,\tau),
&\varth_4(z,\tau):=\varth_{01}(z,\tau).&
\end{align}
The Dedekind eta function is defined as
\[
\eta(\tau):=q^{1/24}\prod_{n=1}^\infty (1-q^n),
\]
where $q=e^{2\pi i\tau}$.
The Eisenstein series are given by
\[
E_{2n}(\tau)
=1+\frac{2}{\zeta(1-2n)}
\sum_{k=1}^{\infty}\frac{k^{2n-1}q^k}{1-q^k}.
\]
The Weierstrass $\wp$-function is defined as
\[
\wp(z|2\omega_1,2\omega_3)
:=\frac{1}{z^2}
+\sum_{(m,n)\in\bbZ^2_{\ne (0,0)}}
\left[\frac{1}{(z-\Omega_{m,n})^2}
  -\frac{1}{{\Omega_{m,n}}^2}\right],
\]
where $\Omega_{m,n}=2m\omega_1 + 2n\omega_3$.

We often abbreviate
$\varth_k(0,\tau),\,\eta(\tau),\,E_{2n}(\tau)$ as
$\varth_k,\,\eta,\,E_{2n}$, respectively.

\section{Taylor expansions of Jacobi theta functions}

In this appendix, we explain how to compute the Taylor expansions
of the Jacobi theta functions. 
Given the expressions (\ref{th1ins}), (\ref{thkinc}),
the problem essentially boils down to
the expansions of functions $\sigma$ and $\sigma_k$
as in (\ref{wsexp}), (\ref{wskexp}).
These expansions can be computed by using
some basic properties of the Weierstrass $\wp$-function.
In the following, we use the abbreviation
\[
\sigma(z):=\sigma(z|2\pi,2\pi\tau),\quad
\sigma_k(z):=\sigma_k(z|2\pi,2\pi\tau),\quad
\wp(z):=\wp(z|2\pi,2\pi\tau).
\]

Recall that the $\wp$-function with period $2\pi,\,2\pi\tau$
satisfies the following identity
\eqb\label{wpdiffeq}
\wp'(z)^2
\Eqn{=}4\wp(z)^3-\frac{E_4}{12}\wp(z)-\frac{E_6}{216}\\[1ex]
\label{wpdifffact}
\Eqn{=}4(\wp(z)-e_1)(\wp(z)-e_2)(\wp(z)-e_3).
\eqe
Here $e_k$ are given in (\ref{e1e2e3}).
They are originally defined as
\[\label{ekdeff}
e_k:=\wp(\omega_k),
\]
where $\omega_k$ are half periods. In the present case
we have
\[
\omega_1=\pi,\qquad
\omega_2=-\pi-\pi\tau,\qquad
\omega_3=\pi\tau.
\]
The $\wp$-function admits an Laurent expansion of the following form
\[\label{wplaurent}
\wp(z)=\frac{1}{z^2}+\sum_{n=1}^\infty c_{n}z^{2n}.
\]
The expansion coefficients are determined by the
recurrence relation
\eqb
c_1\Eqn{=}\frac{E_4}{240},\qquad c_2=\frac{E_6}{6048},\\
c_n\Eqn{=}\frac{3}{(n-2)(2n+3)}\sum_{k=1}^{n-2}c_k c_{n-k-1}
 \qquad (n\ge 3).
\eqe
This recurrence relation can be easily derived
by substituting (\ref{wplaurent}) into the identity
\[\label{ddwp}
\wp''(z)=6\wp(z)^2-\frac{E_4}{24},
\]
which follows from (\ref{wpdiffeq}).
Using identities (\ref{wpdiffeq}), (\ref{ddwp}) recursively,
one can express higher derivatives $\wp^{(2n)}(z)$
as polynomials in $\wp(z),E_4,E_6$.
Similarly,
$\wp^{(2n-1)}(z)$ can be expressed as polynomials in
$\wp(z),E_4,E_6$ multiplied by $\wp'(z)$.
In particular, it follows from (\ref{wpdifffact}), (\ref{ekdeff})
that $\wp^{(2n)}(\omega_k)$ are
polynomials in $e_k,E_4,E_6$,
while $\wp^{(2n-1)}(\omega_k)=0$.

We are now in a position to consider
the expansions of $\sigma(z)$ and $\sigma_k(z)$.
They are related to $\wp(z)$ by
\eqb
\label{wswprel}
\partial_z^2\ln\sigma(z)\Eqn{=}-\wp(z),\\
\label{wskwprel}
\partial_z^2\ln\sigma_k(z)\Eqn{=}-\wp(z+\omega_k).
\eqe
Integrating twice the series expansions of the r.h.s.~of
these equations, one obtains
\eqb
\sigma(z)\Eqn{=}z\exp\left(-\sum_{n=1}^\infty
  \frac{c_n}{(2n+1)(2n+2)}z^{2n+2}\right),\\
\sigma_k(z)\Eqn{=}\exp\left(-\sum_{n=1}^\infty
  \frac{\wp^{(2n-2)}(\omega_k)}{(2n)!}z^{2n}\right),
\eqe
where integration constants were chosen accordingly.
These exponential forms are actually
convenient for the computation of the prepotential.
As we explained above,
$c_n$ are polynomials in $E_4,E_6$,
while $\wp^{(2n-2)}(\omega_k)$ are polynomials
in $E_4,E_6,e_k$.

\section{Proof of combinatorial identity}

In this appendix we prove the identity (\ref{keyidentity}).
The l.h.s.~of (\ref{keyidentity}) reads
\eqb
\lefteqn{\mbox{l.h.s.}}\nn\\
\Eqn{=}
\sum_{k,l=1}^N\sum_{(i,j)\in R_k}
 \left[\left(\mu_{k,i}+\mut_{l,j}-i-j+1\right)^2-(j-i)^2\right]\nn\\
\Eqn{=}
 \sum_{k,l=1}^N\sum_{i=1}^{\ell(R_k)}\sum_{j=1}^{\mu_{k,i}}
 \left[\left(\mu_{k,i}-j+1-i\right)^2
  +2\left(\mu_{k,i}-j+1-i\right)\mut_{l,j}
  +\left(\mut_{l,j}\right)^2-(j-i)^2\right].\nn\\
\eqe
By introducing a new index $\tj:=\mu_{k,i}-j+1$,
the sum over $j=1,\ldots,\mu_{k,i}$ in the first term
can be rewritten as the sum of $(\tj-i)^2$
over $\tj=1,\ldots,\mu_{k,i}$.
We then see that the first term and the last term cancel each other.
Thus we have
\eqb
\mbox{l.h.s.}
\Eqn{=}
\sum_{k,l=1}^N\sum_{(i,j)\in R_k}
 \left[2\left(\mu_{k,i}-j+1-i\right)\mut_{l,j}
  +\left(\mut_{l,j}\right)^2\right]\nn\\
\label{lhs2}
\Eqn{=}
\sum_{k,l=1}^N\sum_{j=1}^{\ell(\Rt_k)}\sum_{i=1}^{\mut_{k,j}}
 \left[\left(2\mu_{k,i}-2j+1\right)\mut_{l,j}
 +(-2i+1)\mut_{l,j}+\left(\mut_{l,j}\right)^2\right].
\eqe
Let us first evaluate the sum of the last two terms.
By performing the sum
over $i=1,\ldots,\mut_{k,j}$, they become
\[
\sum_{k,l=1}^N\sum_{j=1}^{\ell(\Rt_k)}
 \left[-\left(\mut_{k,j}\right)^2\mut_{l,j}
  +\mut_{k,j}\left(\mut_{l,j}\right)^2\right].
\]
Since $\mut_{k,j}$ is defined as
$\mut_{k,j}=0$ for $j>\ell(\Rt_k)$,
the sum over $j$ in the above expression can be replaced by
that over $j=1,\ldots,\infty$.
It then becomes clear that the expression
in the sum over $k,l$ is actually antisymmetric
under the exchange of $k$ and $l$.
Therefore the sum vanishes.

Now, we are left with
\[
\mbox{l.h.s.}
=
\sum_{k,l=1}^N\sum_{j=1}^{\ell(\Rt_k)}\sum_{i=1}^{\mut_{k,j}}
 \left(2\mu_{k,i}-2j+1\right)\mut_{l,j}\,.
\]
It is easy to see that the following relation holds
\[
\sum_{i=1}^{\mut_{k,j}}\left(\mu_{k,i}-j\right)
=\sum_{j'=j+1}^{\ell(\Rt_k)}\mut_{k,j'}\,.
\]
Using this relation, we finally obtain
\eqb
\mbox{l.h.s.}
\Eqn{=}
\sum_{k,l=1}^N\sum_{j=1}^{\ell(\Rt_k)}
 \left(2\sum_{j'=j+1}^{\ell(\Rt_k)}\mut_{k,j'}+\mut_{k,j}\right)
 \mut_{l,j}\nn\\
\Eqn{=}
\sum_{k,l=1}^N\sum_{j=1}^{\infty}
 \left(2\sum_{j'=j+1}^{\infty}\mut_{k,j'}+\mut_{k,j}\right)
 \mut_{l,j}\nn\\
\Eqn{=}
\sum_{k,l=1}^N\sum_{j=1}^{\infty}
 \left(\sum_{j'=1}^{\infty}\mut_{k,j'}
      -\sum_{j'=1}^{j-1}\mut_{k,j'}
      +\sum_{j'=j+1}^{\infty}\mut_{k,j'}
 \right)
 \mut_{l,j}\nn\\
\Eqn{=}
\sum_{k,l=1}^N
 \left(\sum_{j,j'=1}^{\infty}\mut_{k,j'}\mut_{l,j}
      -\sum_{0<j'<j}\mut_{k,j'}\mut_{l,j}
      +\sum_{0<j<j'}\mut_{k,j'}\mut_{l,j}
 \right)\nn\\
\Eqn{=}
\sum_{k,l=1}^N|R_k||R_l|
=\mbox{r.h.s.}
\eqe
%


\renewcommand{\section}{\subsection}
\renewcommand{\refname}

\end{document}